\documentstyle[aps,amsfonts,epsfig]{revtex}
\begin{document}
\title{Spin twists, domain walls,\\
and the cluster spin-glass phase of weakly doped cuprates}
\author{K.S.D. Beach\cite{byline} and R.J. Gooding}
\address{Dept. of Physics, Queen's University, \\ 
Kingston, Ontario, Canada K7L 3N6}
\date{\today}
\maketitle

\begin{abstract}

We examine the role of spin twists in the formation of domain
walls, often called stripes, by focusing on the spin textures
found in the cluster spin glass phases of ${\rm La_{2-x}Sr_xCuO_4}$
and ${\rm Y_{1-x}Ca_xBa_2Cu_3O_6}$.
To this end, we derive an analytic expression for the
spin distortions produced by a frustrating bond, both
near the core region of the bond and in the far field,
and then derive an expression for interaction energies
between such bonds. We critique our analytical theory
by comparison to numerical solutions of this problem
and find excellent agreement. By looking at
collections of small numbers of such bonds localized
in some region of a lattice, we demonstrate the stability
of small ``clusters'' of spins, each cluster having its
own orientation of its antiferromagnetic order parameter. 
Then, we display a domain wall corresponding to spin twists 
between clusters of locally ordered spins showing how spin twists 
can serve as a mechanism for stripe formation. Since the charges 
are localized in this model, we emphasize that these domain walls 
are produced in a situation for which no kinetic energy is present
in the problem.

\end{abstract}

\section{Introduction:}

The now frequent experimental observations of spin and/or charge 
modulations in the cuprate superconductors and related doped transition 
metal oxides \cite{jtran review} was predicted by the ``frustrated phase 
separation'' phenomenology of Emery and Kivelson.  Their 
theoretical considerations \cite {ek1} involved the assertion that a 
doped Mott insulator phase separates as a consequence of the competition 
between the kinetic energy of mobile holes and the magnetic energies of 
an antiferromagnetic (AFM) phase with long-range correlations. While it 
seems unlikely that this scenario is correct in the strong correlation 
limit \cite {whitedec99}, when one adds the Coulomb interaction into this 
problem Emery and Kivelson argued that macroscopic phase separation 
became ``frustrated'', and the resulting anomalous normal state possessed 
low-energy fluctuations corresponding to stripes, or domain walls 
\cite {fps}.  To connote that these entities correspond to metallic 
stripes, it is now common to refer to these structures as ``rivers of 
charge'' \cite{ames}.

Recent neutron scattering studies \cite{kyamada} of the single-layer 
La$_{2-x}$Sr$_x$CuO$_4$ (LSCO) system have revealed that (at least in 
experimental results to date) an elastic magnetic response associated 
with static stripe-like correlations are only found in (i) low $x$ 
systems ($x \lesssim 0.06$) at low temperatures such that the transport 
is that of a doped semiconductor \cite{lai}, and (ii) in the famous 
$x \approx 1/8$ system. One could interpret both the strongly disordered, 
low temperature results, and the $x \approx 1/8$ data, as evidence that 
pinning effects are necessary, either from disorder or commensurability 
interactions, to produce static stripes. Such arguments are consistent 
with the successful approach of Tranquada and coworkers in producing 
static stripes that could then be observed in scattering experiments in 
a variety of systems \cite{jtran review}.

The low $x$, low temperature region of the LSCO phase diagram in which
the static magnetic stripe correlations are found corresponds to the 
spin-glass phase of LSCO. Early magnetic resonance work on this system
\cite{cho} suggested that it is appropriate to think of this phase as a 
cluster spin glass, so named because small clusters of spins achieve 
their own short-range AFM correlations, but the cluster-cluster ordering 
is spin-glass like. A numerical simulation of this phase, coupled with 
new crystals and new susceptibility data, lended support to this 
characterization \cite{skyrms2}.

In the latter paper \cite{skyrms2}, one conundrum associated with the 
{\it mechanism} behind the formation of the cluster spin glass phase was 
pointed out, and goes as follows: The frustrated phase separation 
phenomenology claims that support for such physics is found in 
the existence of the cluster spin glass phase \cite {fps}. 
However, detailed analysis of the transport in this region of the LSCO 
phase diagram concludes that the transport is similar to that of a doped 
semiconductor \cite{lai}. Thus, at least in the low-temperature cluster 
spin-glass phase the competition between kinetic and magnetic energies 
does not exist in the form proposed originally by Emery and Kivelson, and 
the question can then be asked, does the absence of the holes' kinetic 
energy from extended states not eviscerate the frustrated phase 
separation phenomenology as a viable mechanism associated with the 
formation of the cluster spin glass phase? Put another way, the numerical 
simulations of Ref. \cite {skyrms2} that found evidence for the character 
of the spin-glass phase being like that of a cluster spin glass produced 
this spin texture with zero kinetic energy, and thus is the
kinetic energy a necessary ingredient in the formation of stripe
phases \cite {conun}? 

This question becomes more important in view of recent experiments of 
Julien, {\it et al.} \cite{mhj}. These NMR/NQR results demonstrated the 
existence of a so-called charge glass at higher temperatures, followed by 
the appearance of the superconducting phase at lower temperatures, 
followed by the cluster spin glass phase at the lowest temperatures. 
The frustrated phase separation phenomenology predicts this sequence of 
charge glass/cluster spin glass phases, and thus unlike the above 
arguments, suggests that the cluster spin glass is stabilized by the 
pinning of the charge stripes by defects, followed by the subsequent 
freezing of the spin degrees of freedom within clusters defined by the 
pinned stripes of the charge glass phase. Unfortunately, again, the 
transport of this system at low temperatures is insulating (for, say, 
$T < 75~K$), so in the spin glass phase there are no rivers of
charge which could ``carve out'' the domain walls of the cluster spin glass
phase!
 
In this manuscript we present analytical results that formalizes
the claim that via quenched disorder from the Sr impurities one can
produce the topology of pinned stripes without any kinetic energy. To this
end, we derive the spin distortion pattern produced by such quenched
disorder (which frustrates the background AFM order), and then demonstrate
that this theory successfully predicts {\em the stability of localized
clusters of AFM correlated spins produced in a situation with zero kinetic
energy}. In this case, the origin of the stripes associated with the 
domain walls comes from the spin twists between the clusters, the 
clusters themselves having been produced by spin twists of the spin 
texture as the background spins attempt to accommodate the frustrating 
magnetic interactions produced by the quenched disorder. A well known 
extrapolation to higher temperatures \cite {ss1} then implies that the
qualitatively identical spin twists generated by mobile carriers must be 
part of the mechanism associated with the formation of stripes.

We wish to make clear that our paper does not claim to be the first to 
propose that magnetic interactions in general, and spin twists
in particular, are important in the formation of stripes. 
Firstly, the work of Salem and one of us \cite {diluteFM} investigated
the problem of frustrating FM bonds whose locations could be chosen
such that the ground-state energy was minimized. It was found that
when quantum fluctuations were included, if the magnitude of the
frustrating interaction was smaller than that of the background
majority spins, periodic stripes of frustrating bonds were the
ground state configuration. (So, in these ground states, again
there is no kinetic energy {\em but} stripe phases are indeed
encountered.) Secondly, when the frustrating bonds cannot chose their 
(static) positions but are fixed by the Sr impurity ions, the
numerical simulations (mentioned above) of Ref. \cite {skyrms2}
suggested the presence of domain walls between the clusters of
the cluster spin-glass phase. More recently, work of Stojkovic
and coworkers \cite {stojk} examined a version
of the mobile hole problem
by implementing a purely magnetic model that included the
long-ranged spin twists produced by mobile holes (as well as the
frustrating Coulombic energy between the carriers) and found
many of the magnetic structures encountered in \cite {diluteFM},
including stripe phases. Lastly, White and Scalapino, 
who find evidence for stripe structures in the 
$t$--$J$ model for mobile holes \cite{whitestripes}, 
note that the charge and spin distributions of striped structures
attempt to accommodate the frustration (read: spin twists) on the 
magnetic background produced by the mobile holes. 

Looking at the totality of the evidence in this and the above-mentioned
papers we believe that one can make a strong case that there is
a {\em similar mechanism} at work in the formation of stripes
in all of these situations, and that this mechanism is spin twists. 

Our paper is organized as follows. In the next section we present a 
detailed theoretical analysis of the effects of quenched disorder on 
the spin texture in systems such as weakly doped LSCO, producing a 
reliable analytical theory of the spin distortions both near the 
frustrating bond and in the far field. We use this distortion field to 
produce an accurate interaction functional between pairs of such bonds, 
which we then use to demonstrate the stability of such clusters in the 
cluster spins glass phase. In particular, this leads to a clear 
identification of the local AFM order parameter of each cluster.
Finally, we show the resulting spin texture between two such clusters,
and demonstrate how (local) stripe configurations can be stabilized in
the cluster spin-glass phase.

\section{Core solution and energies of the single bond problem:}

\subsection{Hamiltonian and definitions}

We consider the familiar model of magnetism in the CuO planes
of the high $T_{\text{c}}$ cuprates in
which the copper ions and oxygen holes are treated as a lattice
of spins governed by a Heisenberg Hamiltonian
\begin{equation} 
    H = - \sum_{\langle ij \rangle} J_{ij} {\bf S}_i 
    \cdot {\bf S}_j 
\label{eq:ham1}
\end{equation}
where $\langle ij \rangle$ denotes a summation over near neighbour
pairs of spins and the $J_{ij}$ are the exchange interaction integrals. 
For an undoped lattice these spins are the Cu spins and the
exchange integrals are equal and negative; the ground state corresponds 
to an AFM ordered state on a square lattice. In what follows we
simplify our considerations by using classical spins, implying that
only the transverse ({\em viz.}, moment reorientation) and not the 
longitudinal ({\em viz.}, moment magnitude) spin-spin interactions 
are included \cite {ss1}.

The simplest model of the effects of doping in weakly doped cuprates at 
low temperatures was proposed by Emery, and corresponds to localizing the
holes on oxygen sites and replacing the AFM Cu-Cu superexchange for
this occupied bond with an effective FM exchange. The phase diagram of 
the multiply doped version of this model was produced by Aharony, 
{\em et~al} \cite {aharony}.
Although detailed transport analysis of this part of the LaSrCuO
phase diagram \cite {lai} has shown that a slightly different model 
\cite {skyrms1,skyrms2} of the localized dopants is required for a 
direct comparison to experiments, the FM bond model (which we shall 
refer to as the frustrating bond model) is more amenable to analytical 
study, for reasons that we shall elaborate on below, and shall be used 
throughout this paper.

Thus, we consider the Hamiltonian of Eq.~(\ref{eq:ham1}) wherein the
$i,j$ label sites of a square lattice (that is, only the Cu spins and
the effective interactions between them are considered) and
the exchange interaction integral between two adjacent sites $i$ and 
$j$ has the form
\begin{equation} 
    J_{ij}  = \left\{ \begin{array}{ll}
                          \lambda J & \mbox{with probability $x/2$} \\
                          -J & \mbox{with probability $1-x/2$}
                      \end{array}
              \right. 
\label{eq:effJ}
\end{equation}
where $J$ and $\lambda$ are positive constants and $\lambda$ 
represents the relative strength of the ferromagnetic and 
antiferromagnetic bonds \cite {doping}. The doping level $x$ could be,
say, either the Sr doping level in ${\rm La_{2-x}Sr_xCuO_4}$ or the Ca
doping level in ${\rm Y_{1-x}Ca_xBa_2Cu_3O_6}$
(noting that Neidermayer {\it et~al.} \cite {Ca123} has 
shown that the phase diagrams in these two systems are identical).
If we now choose a coordinate system such that linear combinations 
of $x$- and $y$-directed unit vectors span the lattice, then the 
Hamiltonian can be written explicitly as
\begin{equation} 
    H = -\frac{1}{2} \sum_{i} \sum_{\hat{a}} J_{i, i+\hat{a}} 
    {\bf S}_i \cdot {\bf S}_{i+\hat{a}} 
\label{eq:ham2}
\end{equation}
where $i$ is summed over all lattice sites and
$\hat{a}$ ranges over $\pm \hat{x}, \pm \hat{y}$.
The equilibrium condition corresponds to that of zero torque from
the local effective field at each lattice site:
\begin{equation} 
    \sum_{\hat{a}} J_{i, i+\hat{a}} {\bf S}_i \times 
    {\bf S}_{i+\hat{a}} = 0 \: . 
\label{eq:torque}
\end{equation}

As is well known, the complication of treating a bipartite lattice
(labelling the two sublattices as A and B sites) can 
be avoided by transforming the physical problem of FM bonds in a 
predominantly AFM background into the mathematically equivalent 
problem of AFM bonds in a FM background.  These two pictures can be 
converted one to the other under the simple transformation 
(AFM $\rightarrow$ FM) given below:
\begin{eqnarray} 
    J_{ij} &\mapsto& -J_{ij} \\ \nonumber
    {\bf S}_{i}  &\mapsto& 
        \left\{ 
            \begin{array}{ll}
                +{\bf S}_{i} & \mbox{for $i \in A$} \\
                -{\bf S}_{i} & \mbox{for $i \in B$}
             \end{array}
        \right. \: .
\label{eq:ham3}
\end{eqnarray}

Lastly, we note that the objects under consideration are classical spins, 
and we set their length to be one, scaling $J$ to be $JS^2$.  Further, the
ground states that we discuss in this paper all correspond
to situations in which the spins lie in some plane, and thus
from now on we restrict our formalism to describe planar spins.
We denote the bulk direction of the spins by ${\bf S}_\infty$, and 
at any lattice site $i$ there exists a spin ${\bf S}_{i}$ 
characterized by the angle $\psi_{i}$ between the spin and the $x$-axis:
\begin{equation} 
    {\bf S}_{i} = \hat{x} \cos \psi_{i} + \hat{y} \sin \psi_{i} 
~\Rightarrow~ {\bf S}_{i} \cdot {\bf S}_{j} 
     = \cos \, (\psi_{i} - \psi_{j}) \: .
\label{eq:classS}
\end{equation}
We choose 
$\psi_{i} = \phi_{i} + \psi_{\infty}$ where 
$\psi_{\infty}$ is taken 
to be the average angle of the spins over the bulk of the material.  
That is 
${\bf S}_{\infty} = \hat{x} \cos \psi_{\infty} + \hat{y} \sin \psi_{\infty}$
so that the angle $\phi_i$, defined according to
$\cos \phi_i = {\bf S}_{i} \cdot {\bf S}_{\infty}$, 
represents the deviation of the spin at site $i$ from the bulk 
direction.  The collection $\{\phi_i\}$ of the spin distortions
at each lattice site constitutes the spin texture on the lattice.

\begin{figure}
\begin{center}
\epsfig{file=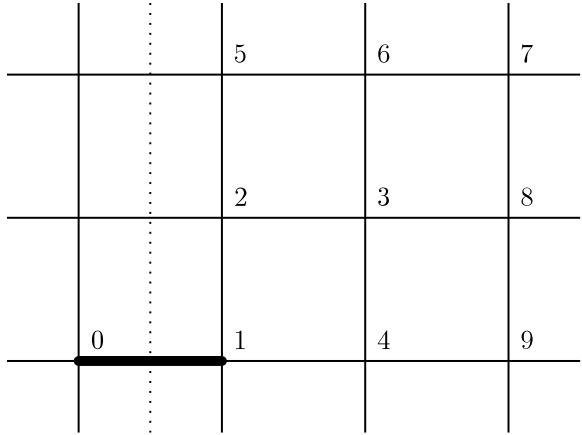,scale=0.75}
\caption{The lattice site numbering scheme.  The frustrating bond sits
between the 0 and 1 sites.}
\label{fig:lattice}
\end{center}
\end{figure}

The numbering scheme for the lattice sites near the frustrating bond
(what from now on we call the bond sites) is shown in 
Fig.~\ref{fig:lattice}.

\subsection{Spin deviations of a frustrating bond}

In a FM lattice doped with a single AFM bond, the ground state 
solution to the spin texture is no longer obvious, and  
such a situation is called frustrated. We wish to produce
an accurate analytical solution to this problem in both
the far-field region and close to the frustrating bond.
This will allow us to accurately track the energies of
both the single and many bond problems.

Consider a purely FM lattice frustrated by the introduction of an
$x$-directed AFM bond between the $(0,0)$ and $(1,0)$ lattice sites.
We shall denote the spin distortions at these sites by $\phi_0$ and
$\phi_1$, respectively.
The Hamiltonian is then written as
\begin{equation}
H  =  - J {\sum_{\langle ij \rangle}}' \cos \, (\phi_{i} - \phi_{j}) 
+ \lambda J \, \cos \, (\phi_0 - \phi_1)
\label{eq:equil_1bond_dis}
\end{equation}
where the prime indicates the omission of the AFM bond from the summation.
It is no longer clear that the trivial solution $\phi_i=0$ represents
the ground state since there may
be another solution with $\pi/2 < |\phi_0 - \phi_1| < 3\pi/2$ 
that the takes the system to a lower energy state.

As is well known, considerable information can be gained from an
examination of the dynamical properties of a linearized system
of equations of motion about a proposed equilibrium structure.
Here, we briefly outline this formalism, since it will be important 
to our later work. For our purposes, the dynamical behaviour of a 
lattice of $n$ independent spins is modeled sufficiently by 
\begin{equation}
\ddot{\phi_i} = - \sum_{\hat{a}} \sin \, (\phi_i - \phi_{i+\hat{a}}) 
\ \ \ i = 1,2, \ldots ,n \: .
\label{eq:dyns}
\end{equation} 
Close to the ordered equilibrium state $\phi_i = 0$, this behaviour is 
governed by the linearized system $\ddot{{\bf x}}  = -M {\bf x}$
where $M = Df({\bf 0})$ is the derivative matrix of $f$ evaluated at 
the origin and ${\bf x}$ is related to the spin texture according to the 
row vector ${\bf x}^T = [ \phi_1 , \phi_2, \phi_3 , \ldots , \phi_n ]$. 

Solving for the normal modes, and taking note of negative eigenvalues,
the instability of the system to a non FM ordered ground state can be 
identified. This technique, when applied for larger and larger systems, 
reproduces the known result \cite {vann} that an instability is first
reached at $\lambda_c = 1$ and that there is only one stable spin texture 
for all $\lambda$ exceeding $\lambda_c$.

Unlike such instability analysis, or the Fourier-based approach of 
Vanninemus, {\it et~al.} \cite {vann}, here we wish to develop a 
continuum theory capable of describing an infinite lattice including 
the spins in the {\em immediate neighbourhood} of the frustrating bond. 
To this end, we proceed as follows. 

Let $\phi$ be a function of a continuous variable ${\bf r}$ which ranges 
over the entire $xy$-plane such that $\phi_i  \mapsto  \phi({\bf r}_i)$.
Then, provided that $\phi$ is a smooth, slowly varying function of 
position, we may approximate the equilibrium condition for
the undoped system (to lowest order \cite {higherorder}) by
\begin{equation} 
\nabla^2 \phi = 0 \: . 
\label{eq:laplace}
\end{equation}

Now consider a FM lattice frustrated by the introduction of a single 
$x$-directed AFM bond. The equilibrium spin distortions away from the 
core of the frustration are governed by Laplace's equation. Choose the 
origin of the coordinate system centred on the bond, and solve Laplace's 
equation, in polar coordinates, by separation of variables. The 
imposition of the appropriate solution symmetries \cite {syms} yields
\begin{equation} 
\phi (r,\theta) = \sum_{m=1,3,5,\dots}^{\infty} r^{-m} A_m \cos m\theta \: 
\label{eq:laplace_soln}
\end{equation} 
where we have adopted the convention that summations are over odd 
indices only.

It is clear that for sufficiently large $r$, the lowest order term 
dominates. That is to say, far from the bond the distortions are dipolar:
\begin{equation} 
\phi({\bf r}) = \frac{{\bf p} \cdot {\bf r}}{r^2} 
\label{eq:dipole1}
\end{equation}
where ${\bf p} = A_1 \hat{x}$ or ${\bf p} = A_1 \hat{y}$.  This agrees 
with the well known results in the literature (see, {\it e.g.}, 
Refs. \cite{vann,aharony}). Unfortunately this result is inadequate for 
our purposes, since we also require the spin distortions near the bond. 
As we show below, a previous attempt \cite {bogdan} fails,
and thus we present new arguments leading to a valid solution close to and
far away from the frustrating bond. Other work has been unable to solve 
analytically this part of the frustrating bond problem \cite {aharonynmr}
(although it is clear that they are aware of the issues that
we have finally solved).

We have written down a general solution to the static spin texture
on the infinite lattice due to a single AFM bond, and that solution
consists of a linear combination of an infinite number of possible
solution modes.  However, since we have shown that the single bond
system has only one stable solution, we expect that any 
prepared state will decay into the state of lowest energy given
by the $m=1$ solution in Eq.~(\ref{eq:laplace_soln}).
That is every $A_m \rightarrow 0$ for $m \neq 1$.
Nonetheless, we run into the difficulty that the field equation 
to which Eq.~(\ref{eq:laplace_soln}) is a solution is not strictly valid 
at the bond sites. Consequently, we cannot expect that these solutions 
will perform well near the bond itself.  
Indeed, we find that the purely dipolar solution with $1/r$ 
fall-off fits numerical solutions extremely well 
$(\sim 0.5 \%)$ up to three or four
lattice sites away from the bond, but that in the core of the frustration
the deviation becomes quite large. At the bond sites themselves, 
the error is $\sim 25\%$.  

We may ask, of course, why such a description does not suffice if, for
the most part, we are interested in the spin distortions away from
the bond.  Surely we can tolerate a small error at a handful of
lattice sites?  The answer is that we cannot.
Since the spin distortions are most severe in the immediate vicinity
of the bond, the spins in the core of the frustration are a large 
contributor to the magnitude of the total energy stored in the 
distortions. Thus a proper calculation of the energy in the system 
requires that we model the core correctly. Equally important is that, 
for a given solution mode, the local equilibrium condition at the bond 
site determines the overall magnitude of the spin distortions. That is, 
it fixes the magnitude of the dipole moment associated with the 
distortion field. Kovalev and Bogdan, who suggested a continuum approach 
for the core region of this problem \cite{bogdan}, fall into precisely 
this trap and hence obtain the wrong magnitude for the long range 
behaviour for the spin distortions (see below). 

It remains to be answered how we might treat the spin distortions
around the bond.  Ideally, we would like to treat the AFM character
of the interaction in the core as a small perturbation on the
field equations, but this is not possible since the continuum
formalism is badly behaved at the origin under
the symmetries we have imposed.
A second possibility would be to derive a separate discrete solution 
valid in the core and to match it smoothly onto the exterior continuum 
solution.  However, we are inclined to avoid such
a patch-work approach.  Not only is it somewhat inelegant, but it also 
defeats the purpose of introducing the continuum formalism, namely, 
to do away with discrete calculations altogether.

Instead, we make use of the fact that the sets
$\{\cos(m\theta)/r^m\}$ and $\{\sin(m\theta)/r^m\}$
are complete (in the sense that any static spin texture satisfying 
the symmetries \cite {syms} specified by a single $x$- or $y$-directed bond
can be expanded in one of these bases). Thus, to solve for the 
spin distortions everywhere on the lattice is essentially to fix the 
values of the coefficients $\{A_m\}$. In the following, we attempt to 
expand the solution to the spin distortions of an $x$-directed AFM bond 
near the origin in the basis $\{\cos(m\theta)/r^m\}$. 

To start, we expect that the coefficient $A_1$ must dominate the 
others since $\cos (\theta)/r$ is the mode of lowest energy. Further, 
convergence at the bond sites requires that $(A_m) \rightarrow 0$ faster 
than $2^{-m}$ as $m \rightarrow \infty$. Thus, it is meaningful to treat 
the expansion $\phi^{(n)}(r,\theta)$, consisting
of the first $n$ terms of Eq.~(\ref{eq:laplace_soln}) as an approximate 
solution. We can then apply the local equilibrium condition at $n$ sites
around the bond to determine the $n$ coefficients.

For concreteness, consider the four term expansion 
\begin{equation} 
\phi(r,\theta) = \phi^{(4)}(r,\theta) 
= A_1 \frac{\cos \theta}{r} + A_3 \frac{\cos 3\theta}{r^3}
 + A_5 \frac{\cos 5\theta}{r^5} + A_7 \frac{\cos 7\theta}{r^7} \:. 
\label{eq:4term1}
\end{equation}
To solve for its four coefficients we require the 
$9 \times 4$ transformation matrix
\begin{equation}
T := {\left[ \frac{\partial \phi_i}{\partial A_j} \right]}_{9 \times 4} 
= \left[
\begin{array}{cccc}
2 &  8   &       32    &       128   \\ \\
\frac{2}{5} & -\frac{88}{125} & \frac{1312}{3125} 
& \frac{3712}{78125} \\ \\
\frac{6}{13} & -\frac{72}{2197} & -\frac{19104}{371293} 
& -\frac{569472}{62748517} \\ \\
\frac{2}{3} & \frac{8}{27} & \frac{32}{243} & \frac{128}{2187} \\ \\
\frac{2}{17} & -\frac{376}{4913} & \frac{35872}{1419857} 
& -\frac{2566016}{410338673} \\ \\
\frac{6}{25} & -\frac{936}{15625} & -\frac{7584}{9765625} 
& \frac{9784704}{6103515625} \\ \\
\frac{10}{41} & -\frac{920}{68921} & -\frac{335200}{115856201} 
& \frac{609920}{194754273881} \\ \\
\frac{10}{29} & \frac{520}{24389} & -\frac{47200}{20511149} 
& -\frac{14926720}{17249876309} \\ \\
\frac{2}{5} & \frac{8}{125} & \frac{32}{3125} & \frac{128}{78125}
\end{array} \right]
\label{eq:Tmatrix}
\end{equation}
(generated using a symbolic algebra computer package) and the matrix
\begin{equation} 
M = \left[ \begin{array}{rrrrrrrrr}
(3-2\lambda) & -2 & 0 & -1 & 0 & 0 & 0 & 0 & 0 \\
-1 &5 &-1 &0 &-1 &0 &0 &0 &0 \\
0&-1&4&-1&0&-1&0&-1&0 \\
-1&0&-2&4&0&0&0&0&-1 \\
 \end{array} \right] 
\label{eq:Mmatrix}
\end{equation}
of linearized equilibrium conditions at sites 1 through 4.
Defining the row vector ${\bf a}^T = [ A_1, A_3, A_5, A_7 ]$
the determination of $\{A_m\}$ is equivalent to solving the
homogeneous system of equations $MT{\bf a} = 0$. 
The requirement that $\det (MT) = 0$
yields a critical value $\lambda_{\text{c}} \, \dot{=} \, 1.0113$ (very
close to the true value $\lambda_{\text{c}}=1$) for which 
\begin{equation} 
{\bf a} =  [ 0.5987, 0.0742, 0.1960, -0.0552 ] \times \phi_1 \:. 
\label{eq:a soln}
\end{equation}
That is to say, the best four term expansion reads
\begin{eqnarray}
\phi^{(4)}(r,\theta) & = & \phi(r,\theta) 
= A_1 \frac{\cos \theta}{r} + A_3 \frac{\cos 3\theta}{r^3}
 + A_5 \frac{\cos 5\theta}{r^5} + A_7 \frac{\cos 7\theta}{r^7} 
\\ \nonumber
& = & \frac{{\bf p} \cdot {\bf r}}{r^2} + \sum_{m=3,5,7} A_m 
\frac{\cos m\theta}{r^m}
\label{eq:4term2}
\end{eqnarray}
with
\begin{equation}
p = |{\bf p}| = A_1 = +0.5987 \phi_1,~
A_3 = +0.0742 \phi_1, 
A_5 = +0.1960 \phi_1,~ A_7 = -0.0552 \phi_1 
\label{eq:4term3}
\end{equation}
The coefficients of $\phi^{(n)}(r,\theta)$ for $n$ being increased from 
1 to 4 are presented in Table~\ref{table1}.

\begin{table}
\caption{Spin distortion amplitudes of $\phi^{(n)}(r,\theta)$ for 
$n=1,2,3,4$.} 
\begin{tabular}{|c|c|c|c|c|c|} \hline
$n$ & $\lambda_{\text{c}}$ & $A_1/\phi_1$ & $A_3/\phi_1$ 
& $A_5/\phi_1$ & $A_7/\phi_1$ \\ \hline
1 & 17/15 & 15/22  & - & - & - \\
2 & 1.0222 & 0.6274 & -0.0318 & - & - \\
3 & 1.1551 & 0.3516 & -0.1220 & 0.0398 & - \\
4 & 1.0113 & 0.5987 & 0.0742 & 0.1960 & -0.0552 \\ \hline 
\end{tabular}
\label{table1}
\end{table}

What this calculation provides that the other does not is the value of 
the multiplicative factor 
$\partial A_1/ \partial \phi_1 = A_1/\phi_1 \sim 0.6$ 
relating the magnitude of the spin distortions at the bond sites to the 
magnitude of the dipole moment associated with the bond itself.

We have shown that the spin distortions are given everywhere by
\begin{equation} 
\phi(r,\theta) = A_1 \frac{\cos \theta}{r} + A_3 \frac{\cos 3\theta}{r^3}
 + A_5 \frac{\cos 5\theta}{r^5} + \cdots 
\label{eq:4term4}
\end{equation}
As we have seen, however, this expression is unwieldy in that it 
requires the application of infinitely many local equilibrium conditions 
to fully determine the coefficients $\{A_m\}$.
Even to calculate the coefficients of a finite series expansion of 
several terms is computationally expensive.  
Ideally, what we would like to have is a solution dependent on a 
single parameter whose value
is determined by applying a single boundary condition at the bond itself.
Here we now outline such a method. We find that a simple assumption
on the distribution of modes can provide this very result.

We proceed by assuming that the spectrum of modes can be modeled by
\begin{equation} 
A_{2n+1} = (-1)^n \frac{2A_1}{(2n+1)2^{2n+1}} 
\label{eq:ansatz1}
\end{equation} 
or, in somewhat simplified notation,
\begin{equation} 
A_k = (\pm) \frac{2A_1}{k 2^k} 
\label{eq:ansatz1a}
\end{equation}
where the index $k$ is taken to be odd and the sign is taken alternately 
positive and negative. This form falls off just fast enough to make the 
series converge --- besides this seemingly naive reason, we appeal to 
its success (described in detail below) to justify its usage.

Under this assumption
\begin{eqnarray}
\phi(r,\theta) & = & A_1 \frac{\cos \theta}{r} 
+ A_3 \frac{\cos 3\theta}{r^3}
+ A_5 \frac{\cos 5\theta}{r^5} + \cdots \\ \nonumber
& = & 2A_1 \left( \frac{\cos \theta}{2r} - \frac{1}{3} 
\frac{\cos 3\theta}{(2r)^3} + 
\frac{1}{5} \frac{\cos 5\theta}{(2r)^5} - \cdots \right) \:. 
\label{eq:4tmerm5}
\end{eqnarray}
Now, the magnitude of the dipole moment in terms of the distortion 
at the bond site follows immediately from solving $\phi(1/2,0)=\phi_1$ 
self-consistently.  We find that $A_1 = \frac{2}{\pi} \phi_1$
and hence
\begin{equation} 
\phi(r,\theta) = \phi_1 \frac{4}{\pi} \sum_k (\pm)\frac{\cos k\theta}{k(2r)^k} 
\label{eq:1bdsoln}
\end{equation}
since 
\begin{equation} 
\phi_1 = \phi(1/2,0) = \phi_1 \frac{4}{\pi} \left( 1 - \frac{1}{3} + \frac{1}{5} - \cdots \right) 
\label{eq:phi1eq}
\end{equation}
which is identically equal to $\phi_1$.
The spin distortions at the remaining sites in the immediate neighbourhood 
of the bond are as follows:
\begin{equation}
\phi_2 = \phi(\sqrt{5}/2, \arctan(2))  =  0.2951672353 \phi_1 ~,~
\phi_4 = \phi(3/2,0)  =  0.4096655294 \phi_1
\label{eq:phi2_4 solns}
\end{equation}
(In fact, one may prove the identity $2\phi_2 + \phi_4 \equiv \phi_1 $, which we shall use later on in
this paper.)

Notice that as $r$ becomes large, we get
\begin{eqnarray} 
\phi(r,\theta) &\rightarrow & \phi_1 \frac{4}{\pi} \cos \theta \arctan \left( \frac{1}{2r} \right)
\rightarrow \phi_1 \frac{4}{\pi} \frac{\cos \theta}{2r} \\ \nonumber
&=& \phi_1 \frac{2}{\pi} \frac{\cos \theta}{r} 
\label{eq:4term5}
\end{eqnarray}
so that the solution retains its familiar long range behaviour.  That is
\begin{equation} 
\phi({\bf r}) = \frac{{\bf p} \cdot {\bf r}}{r^2} 
\label{eq:4term asym}
\end{equation}
but now with
\begin{equation} 
p = |{\bf p}| = \frac{2}{\pi} \phi_1 \: . 
\label{eq:new dipole}
\end{equation}

\begin{figure}
\begin{center}
\epsfig{file=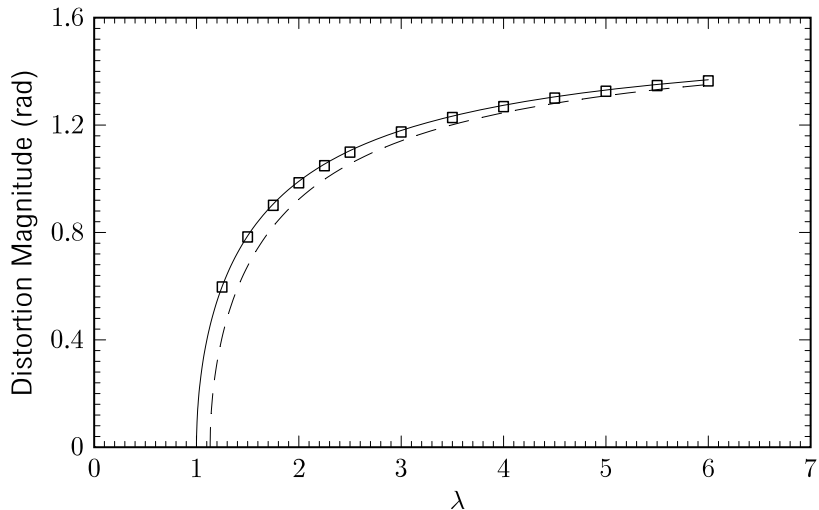,scale=0.75}
\caption{The distortion magnitude as a function of
the frustrating AFM bond strength $\lambda$.  The solid line represents
the prediction of our analytical theory, and the broken line is the
prediction based on the solution of Kovalev and Bogdan 
\protect \cite{bogdan}. Open squares are numerically generated data 
points from large lattices (from $40 \times 39$  up to $80 \times 79$ 
lattices) with converged numerical solutions for this quantity.}
\label{fig:dist_vs_lambda}
\end{center}
\end{figure}

What remains is to determine the parameter $\phi_1$.
As promised, the bond furnishes a single boundary condition in the form of
the equilibrium condition applied at either of the bond sites:
\begin{equation} 
-\lambda \sin (2\phi_1) + 2 \sin (\phi_1-\phi_2) + \sin (\phi_1-\phi_4) 
= 0 \:. 
\label{eq:bdy phi1}
\end{equation}
This is an implicit equation for $\phi_1$, and thus for all of the 
spin distortions as a function of $\lambda$. Its solution is plotted in 
Fig.~\ref{fig:dist_vs_lambda}.

Moreover, the linearized equation gives
\begin{equation} 
-\lambda_{\text{c}} 2\phi_1 + 2(\phi_1-\phi_2) + (\phi_1 - \phi_4) = 0 
\label{eq:lin bdy phi1}
\end{equation}
which can be solved explicitly:
\begin{equation} 
\lambda_{\text{c}} = \frac{3}{2} - \frac{2\phi_2 + \phi_4}{2 \phi_1} \equiv 1 \: . 
\label{eq:lin bdy crit}
\end{equation}
That is, our ansatz correctly reproduces the exact critical value of 
$\lambda_c$ !

\begin{figure}
\begin{center}
\epsfig{file=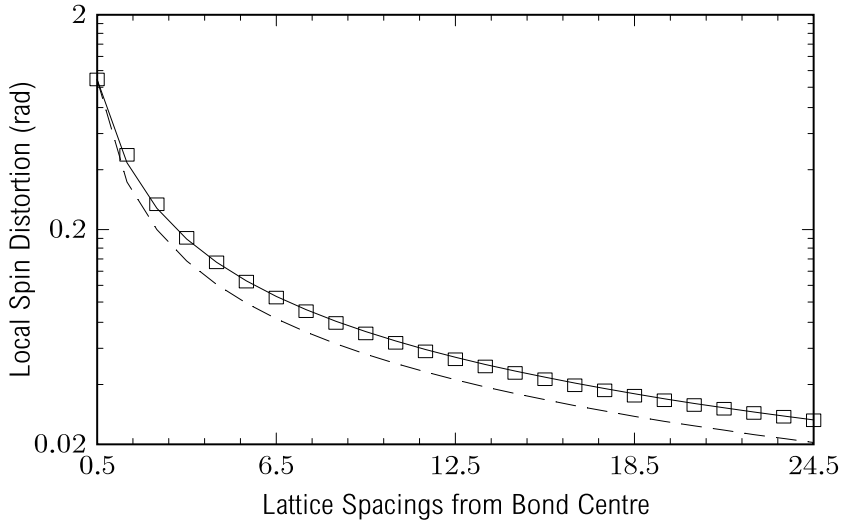,scale=0.75}
\caption{The local spin distortions as a function of distance along 
the $x$-axis for an $x$-directed AFM bond. The solid line represents the 
function $\phi(r,0)$ described in the text.
Open squares are converged numerically generated data points from 
simulations performed on the same size lattices as in 
Fig. 2. The broken line is the $\phi^{(1)}(r,\theta)$ solution 
of Kovalev and Bogdan \protect \cite{bogdan} which fails to predict
quantitatively the correct dipole moment of the long-range distortions.}
\label{fig:dist_vs_r}
\end{center}
\end{figure}

A comparison of the solution Eq.~(\ref{eq:1bdsoln}) 
to numerical simulations is presented in Fig.~\ref{fig:dist_vs_r}; 
clearly, the agreement is excellent, providing the most direct support 
for our ansatz. 

\subsection{Energy functional}

We now calculate the total energy stored in the spin distortions induced
by a single AFM bond.  That these distortions are both small and slowly 
varying in position away from the core allows us to convert the sum of 
the energy contributions into an integral of the energy density 
$(\nabla \phi)^2$. A complete derivation is provided in Appendix~A, 
and we summarize the results below.

To begin, the Hamiltonian is approximated to second order everywhere
except across the AFM bond itself:
\begin{equation}
H 
 \approx  - \sum_{\langle ij \rangle} J_{ij} + \frac{1}{2} 
J {\sum_{\langle ij \rangle}}' (\phi_i - \phi_j)^2 
               - 2 \lambda J \sin^2 \, \phi_1
\label{eq:ham4}
\end{equation}
Of course, the term $-\sum J_{ij}$ represents the total energy of the 
system in the absence of spin distortions so that the energy from the 
distortions alone is given by the latter two terms.
They, in turn, can be expanded using the continuum approximation
\begin{eqnarray}
E_{\mbox{\scriptsize dist}} 
        & = & \frac{1}{4} J \sum_{i \ne 0,1} \sum_{\hat a} 
(\phi_i - \phi_{i+\hat{a}})^2 
  + \frac{1}{2} J \Bigl( 2 (\phi_1 - \phi_4 )^2 + 4 (\phi_1 - \phi_2 )^2 \Bigr) 
     - 2 \lambda J \sin^2 \phi_1 \nonumber \\ 
  & \approx & J \biggl\{ \frac{1}{2} \int_M (\nabla \phi)^2 d^2r
+ (\phi_1 - \phi_4 )^2+2(\phi_1 - \phi_2 )^2 - 2 \lambda \sin^2 \phi_1 \biggr\}
\label{eq:nrgd2}
\end{eqnarray}
where $M$ is the $xy$-plane excluding a small region about the bond centre.

\begin{figure}
\begin{center}
\epsfig{file=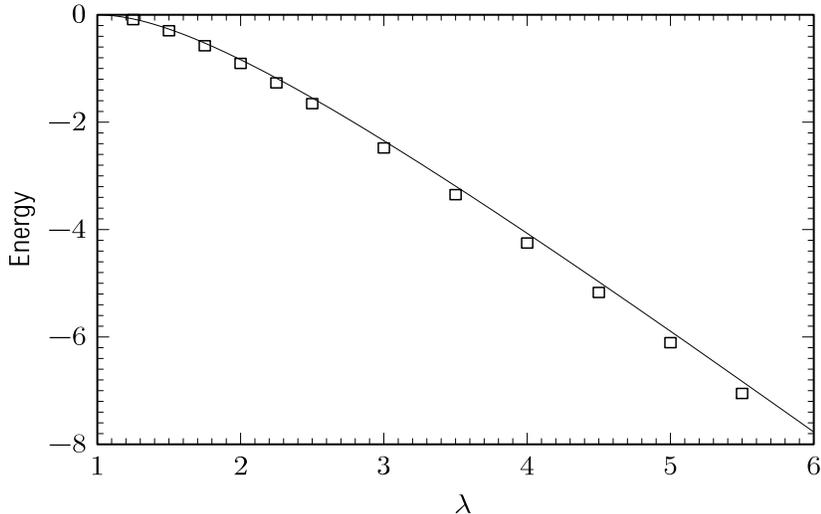,scale=0.75}
\caption{Energy of a single bond (in units of $J$) as a function of
the relative AFM bond strength $\lambda$.  The solid line represents
the prediction of theory, and 
open squares are converged numerically generated data points from
simulations performed on the same size lattices as in 
Fig. 2.}
\label{fig:nrg_vs_lambda}
\end{center}
\end{figure}

An explicit evaluation of the energy using the solution of the previous 
subsection is presented in Appendix~A, wherein the full effect of the 
core region is accounted for. We find
\begin{equation} 
E_{\mbox{\scriptsize dist}} = 2J \left( 
\phi_1^2 - \lambda \sin^2 \phi_1 \right) \: . 
\label{eq:nrgd3}
\end{equation}
This result is compared to numerical simulations in 
Fig.~\ref{fig:nrg_vs_lambda}, 
and again, excellent agreement between our numerical solutions and our 
analytical work is found.

So, now we carry on to the examination of the many-bond problem, 
having an excellent solution to both the core and far-field distortion 
patterns of the single-bond problem, as well as an accurate energy 
functional for an isolated frustrating bond.

\section{Interacting Frustrating Bonds:}

After dealing with a single bond in isolation, the next step toward 
treating a non-zero density of bonds is to determine how bonds interact 
with one another. In this subsection we consider the problem of two bonds.

\begin{figure}
\begin{center}
\epsfig{file=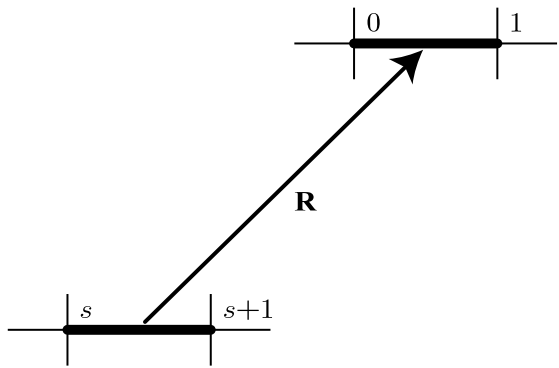,scale=0.75}
\caption{The geometry and labelling used for two parallel bonds separated 
by a vector ${\bf R}$.}
\label{fig:bond_geom}
\end{center}
\end{figure}

Suppose that there is an AFM bond, call it $A$, between the 0 and 1 sites.
Then suppose that another similarly directed 
bond, call it $B$, is placed between the $s$ and $s+1$
sites and that a vector ${\bf R}$ making an angle $\Phi$ with
the $x$-axis connects the two bond centres, as in 
Fig.~\ref{fig:bond_geom}.  

We expect that the energy can be parametrized by two variables
\begin{equation}
\alpha  =  \frac{\phi_1 - \phi_0}{2}~~~~~~~~ 
\beta  =  \frac{\phi_{s+1} - \phi_s}{2}
\label{eq:2vars}
\end{equation}
and that it can be written in the form
\begin{eqnarray}
E_2(\alpha,\beta) & = & E(\alpha) + E(\beta) + Jg\alpha\beta 
\nonumber \\
& = & 2J(1-\lambda)(\alpha^2+\beta^2) 
+  \left(\frac{2}{3} - \delta \right) J \lambda (\alpha^4 + \beta^4) 
+ Jg\alpha\beta
\label{eq:nrg2vars}
\end{eqnarray}
where $g=g({\bf R})$ is a real-valued function of the separation and 
relative orientation of the bonds and $\delta$ is a small correction representing the 
$4^{\mbox{\scriptsize th}}$ order contribution to energy from the spin distortions
which were neglected in Eq.~(\ref{eq:nrgd2}).

The requirement that the spin texture remain invariant up to a global 
sign change under interchange of $\alpha$ and $\beta$ reduces the two 
parameter energy expression (\ref{eq:nrg2vars})
to one of two one parameter expressions $E_2^+$ or $E_2^-$
corresponding to the symmetric ($\alpha = \beta$) and the 
antisymmetric ($\alpha = -\beta$) state.

\emph{Case 1}: $\phi = \alpha = -\beta$.  In this case, the dipoles 
associated with the bond are anti-aligned. The total energy is
\begin{equation}
E_2^-(\phi) = E_2(\phi,-\phi) 
 =  4J  \left( (1-\lambda-\frac{1}{4}g)\phi^2 
+ \frac{1}{2}\left(\frac{2}{3} - \delta \right) \lambda \phi^4 \right) 
~~ < ~~ 2E(\phi) \:.
\label{eq:nrg-neg}
\end{equation}
Minimization with respect to $\phi$ gives
\begin{equation} 
\phi = \sqrt{ (\lambda + \frac{1}{4}g - 1) / 
\left(\frac{2}{3}-\delta\right)\lambda } \:. 
\label{eq:dist-neg}
\end{equation}
This implies that the critical value of $\lambda$ at which the canted 
ground state
first appears is lower than it is for a single AFM bond $\lambda_c = 1$. 

\emph{Case 2}: $\phi = \alpha = \beta$.  This represents a higher energy 
metastable state characterized by aligned dipoles. The energy for this 
configuration is 
\begin{equation}
E_2^+(\phi) = E_2(\phi,\phi) 
= 4J \left( (1-\lambda+\frac{1}{4}g)\phi^2 + 
\frac{1}{2}\left(\frac{2}{3} - \delta \right)
\lambda \phi^4 \right)~~ >~~ 2E(\phi)
\label{eq:nrg-pos}
\end{equation}
with distortion magnitude
\begin{equation} 
\phi = \sqrt{ (\lambda - \frac{1}{4}g - 1) / 
\left(\frac{2}{3}-\delta\right)\lambda } \:.
\label{eq:dist-pos}
\end{equation}
Now the critical value of the coupling constant required for a distorted 
ground state exceeds $\lambda_c = 1$.

Case 1 is of particular interest since it yields the ground state energy 
$E_2^0$ of the two bond system. Further, by re-expressing that energy in 
terms of the energy of a single bond (extracted from Eq.~(\ref{eq:nrgd3}) 
in the limit $\lambda \rightarrow \lambda_c + 0^+$) we can determine the 
energy of interaction between the two bonds.
\begin{equation}
E^0_2  =  -4J \frac{(\lambda + \frac{1}{4}g -1)^2}{\lambda(2/3-\delta)}
=  2E^0 - J\frac{(\lambda-1)}{\lambda(2/3-\delta)}g({\bf R}) - 
J\frac{1}{8\lambda(2/3-\delta)}g^2({\bf R})
\end{equation}
In general, since we expect $g$ to be small, we can write
\begin{equation} 
E_{\mbox{\scriptsize int}} = - 
J\frac{(\lambda-1)}{\lambda(2/3-\delta)}g({\bf R}) \:. 
\end{equation}
However, as $\lambda \rightarrow 1$, we obtain
\begin{equation} 
E_{\mbox{\scriptsize int}} \rightarrow - 
J\frac{1}{8\lambda(2/3-\delta)}g^2({\bf R}), 
\end{equation}
a weak, long-range interaction with a higher power law, a consequence of 
the fact that
the dipolar distortions do not pre-exist in the unperturbed medium at 
$\lambda = 1$.  Such an interaction is analogous to the Van der Waals 
interaction between thermally fluctuating dipoles.

The function $g$ expresses the functional dependence of the interaction
energy on the geometrical configuration of the bonds. We have yet to 
determine its exact form.  All we can say now is that 
\begin{equation} 
\lim_{R \rightarrow \infty} g({\bf R}) = 0 
\end{equation}
which simply formalizes our expectation that two bonds must be 
non-interacting at infinite separation.

The long range spin distortions arising from each of the bonds 
$A$ and $B$ with dipole moments ${\bf p}^A$ and ${\bf p}^B$ is given by
\begin{equation} 
\phi^A({\bf r}) = \frac{{\bf p}^A\cdot {\bf r}}{r^2} \ ~~~\mbox{and}~~~ \ 
\phi^B({\bf r}) = \frac{{\bf p}^B\cdot {\bf r}}{r^2} \:. 
\end{equation}
The linearity of the Laplacian implies that the field equations
admit a superposition principle.  Therefore, we take the total spin 
distortion at each point to be
\begin{equation} 
\phi = \phi^A + \phi^B 
\end{equation}
where $\phi^A$ and $\phi^B$ are the spin distortions from each bond 
in the absence of the other.
  
The energy in the distortions away from the cores goes as
\begin{eqnarray}
\int_{M2} (\nabla \phi)^2 d^2r 
& = & \int_{M2} \bigl(\nabla (\phi^A+\phi^B)^2 \bigr) d^2r \\ \nonumber
& = & \int_{M2} (\nabla \phi^A)^2 d^2r + \int_{M2} (\nabla \phi^B)^2 d^2r 
\nonumber \\
& & +\int_{M2} \nabla \phi^A \cdot \nabla \phi^B d^2r \nonumber
\end{eqnarray}
where $M2 \left( = {\Bbb R} \backslash D_{\epsilon}({\bf 0}) 
\backslash D_{\epsilon}({\bf R}) \right)$ is the $xy$-plane with
disks removed about the bond centres.  The last  term in this expression 
vanishes identically, which indicates that the core of each bond 
interacts with the long-range spin 
distortions of the other and justifies a rather involved calculation 
of the interaction energy which we have relegated to Appendix~B.

What we find is that $E_{\mbox{\scriptsize int}}$ has the form of a 
magnetic dipole interaction. 
Further, although in the preceding discussion we considered only parallel 
bonds,
it is simple to show that these results hold more generally.
Thus, for two bonds which are parallel or perpendicular we have
\begin{equation} 
E_{\mbox{\scriptsize int}} = 
J\frac{2\pi}{R^2} \left\{2({\bf p}^A\cdot\hat{{\bf R}})
({\bf p}^B\cdot\hat{{\bf R}})
- {\bf p}^A \cdot {\bf p}^B \right\} \:. 
\end{equation}

Finally, we can work backwards to find $g({\bf R})$.
For parallel bonds,
\begin{equation}
E_{\mbox{\scriptsize int}}  =  J\frac{2\pi}{R^2} 
\left\{2({\bf p}^A\cdot\hat{{\bf R}})({\bf p}^B\cdot\hat{{\bf R}})
- {\bf p}^A \cdot {\bf p}^B \right\} 
 =  \pm J \frac{(\lambda-1)}{\lambda(2/3-\delta)} 
\frac{8}{\pi R^2}\cos 2\Phi \:.
\end{equation}
That is 
\begin{equation} 
g({\bf R}) = \frac{8}{\pi R^2} \cos 2\Phi \:. 
\end{equation}
We note that the identical calculation for two perpendicular bonds gives
\begin{equation} 
g({\bf R}) = \frac{8}{\pi R^2} \sin 2\Phi \:. 
\end{equation}

\begin{figure}
\begin{center}
\epsfig{file=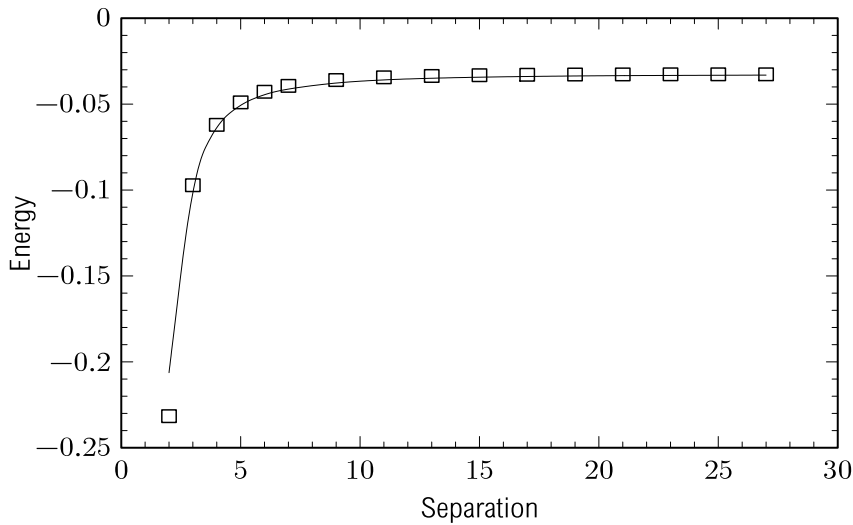,scale=0.75}
\caption{The ground state energy of two parallel bonds, for $\lambda=1.1$, 
at various separations along the $x$-axis, where the solid line is
our prediction and the open squares are converged numerically generated 
data points from simulations performed on the same size lattices as in
Fig. 2.}
\label{fig:nrg_vs_r_1pt1}
\end{center}
\end{figure}

\begin{figure}
\begin{center}
\epsfig{file=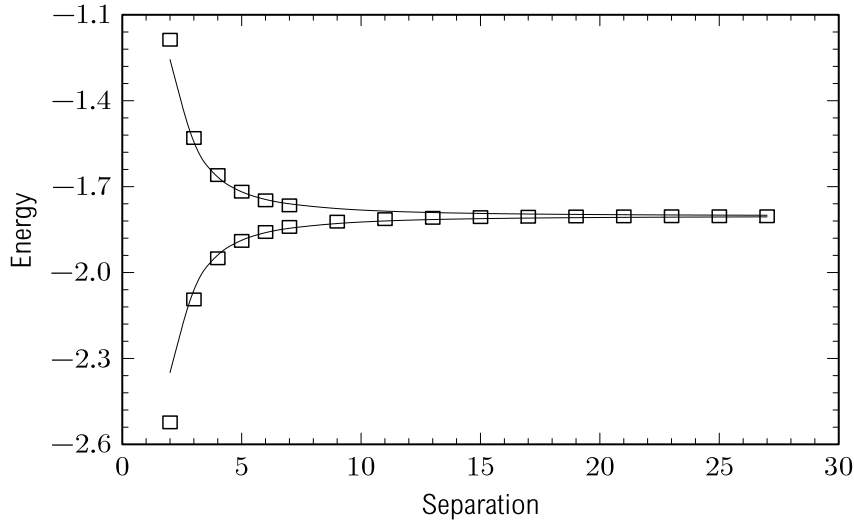,scale=0.75}
\caption{Energy of two parallel bonds, $\lambda=2$, at various 
separations along the $x$-axis, where the solid lines
and open squares are converged numerically generated
data points from simulation performed on the same size lattices
as in Fig. 2. The upper curve corresponds to the 
metastable state with dipoles similarly directed $(\rightarrow 
\rightarrow)$, while the lower energy curve corresponds to oppositely 
directed dipoles ($\rightarrow \leftarrow)$.}
\label{fig:nrg_vs_r_2}
\end{center}
\end{figure}

\section{Clusters in the cluster spin glass phase:}

We should ask whether the results we have obtained so far 
for the one and two bond problems can be generalized to allow us to 
tackle the problem of a lattice frustrated by the presence of any number 
of arbitrarily placed AFM bonds. It should be clear that, in general, 
even for relatively few bonds, the induced spin distortions will be 
very complicated and the energy surface characterized by many closely 
spaced, low-lying states. In such a case one must resort to 
sophisticated computer algorithms to numerically 
generate the ground state spin texture, and the qualitative analysis
of the cluster spin-glass phase from such work has been analyzed
elsewhere \cite{grempel,skyrms2}.

In contrast to that work, here we note that there are configurations 
of suitably high symmetry for which we can confidently treat the spins 
as planar and even solve analytically for the spin distortions and
energies of all the possible states.  The smallest such configuration
is the square {\em cluster} of parallel bonds.

By a cluster we imply a collection of bonds arranged in some local
region on the lattice.  That collection, call it $C$, 
can be thought of as a set of dipole--position pairs:
$\{({\bf p}^{\alpha},{\bf r}^{\alpha})\}_{\alpha \in C} \: . $
Given a high symmetry cluster, for which a spin-planar ground state is 
justified, the spin distortions away from the cores of the bonds are 
given by
\begin{equation} 
\phi({\bf r}) = \sum_{\alpha \in C} 
\frac{{\bf p}^{\alpha} \cdot ({\bf r} 
- {\bf r}^\alpha)}{|{\bf r}-{\bf r}^{\alpha}|^2} 
\end{equation}
which is the solution (unique for the required symmetry) to the equation
\begin{equation} 
\nabla^2 \phi({\bf r}) = - 2\pi \nabla \cdot \sum_{\alpha \in C} 
{\bf p}^\alpha \delta({\bf r}-{\bf r}^\alpha) \:. 
\end{equation}
The total interaction energy of the cluster can be written as the sum 
of all pairwise interactions
\begin{equation} 
E_{\mbox{\scriptsize int}} = \sum_{\alpha<\beta \in C}
J \frac{2\pi}{|{\bf r}^\alpha-{\bf r}^\beta|^2} \big(
2\frac{{\bf p}^{\alpha} \cdot ({\bf r}^\alpha - {\bf r}^\beta) \,
{\bf p}^{\beta} \cdot ({\bf r}^\alpha 
- {\bf r}^\beta)}{|{\bf r}^\alpha-{\bf r}^\beta|^2}
- {\bf p}^{\alpha} \cdot {\bf p}^{\beta} \big)  \:. 
\end{equation}

Thus, for instance, the $L \times L$ square cluster of parallel bonds 
given by
\begin{equation}
{\bf r}^1  =  (0,0),~~
{\bf r}^2  =  (L,0),~~
{\bf r}^3  =  (0,L),~~
{\bf r}^4  =  (L,L),~~
{\bf p}^{\alpha}  =  \pm p\hat{x} \ \ {\rm for}~~\alpha \in C=\{1,2,3,4\},
\end{equation}
has a rather simple interaction energy.  There
are four possibilities depending on the the orientation of each dipole:
\begin{equation} 
E_{\mbox{\scriptsize int}} = -J \frac{8\pi}{L^2}p^2 \,,\, 0 \,,\, 0 \,,\, 
+J\frac{8\pi}{L^2}p^2 
\end{equation}
In practice, however, the degeneracy of the middle two states is lifted 
by higher order terms in the interaction energy. Indeed, the splitting 
observed in numerical simulations enables us to list the four 
distinguished states in order of decreasing energy.
\begin{equation} 
\left| \begin{array}{cc} \rightarrow~~ \rightarrow \\ \leftarrow 
~~\leftarrow \end{array} \right| ~~,~~ 
\left| \begin{array}{cc} \rightarrow~~ \rightarrow \\ \rightarrow 
~~\leftarrow \end{array} \right| ~~,~~
\left| \begin{array}{cc} \rightarrow ~~\rightarrow \\ \rightarrow 
~~\rightarrow \end{array} \right| 
 =  \mbox{first excited state} ~~,~~
\left| \begin{array}{cc} \leftarrow~~ \rightarrow \\ \leftarrow 
~~\rightarrow \end{array} \right|  = \mbox{ground state} 
\end{equation}

The first excited state consists of four similarly directed dipoles.
Destructive interference inside the cluster gives zero net distortion, but 
outside the cluster the dipoles add constructively so that the cluster 
acts like a single unit with a much stronger moment. This results in 
very strong long range spin distortions. Far enough from the bond, the 
spin distortions are given by
\begin{equation} 
\phi({\bf r}) = \frac{4{\bf p} \cdot {\bf r}}{r^2} \:. 
\end{equation}
Since the long range spin distortions mediate the interaction between
bonds, we expect a cluster of this kind to strongly couple to other
bonds in the lattice.

\begin{figure}
\begin{center}
\epsfig{file=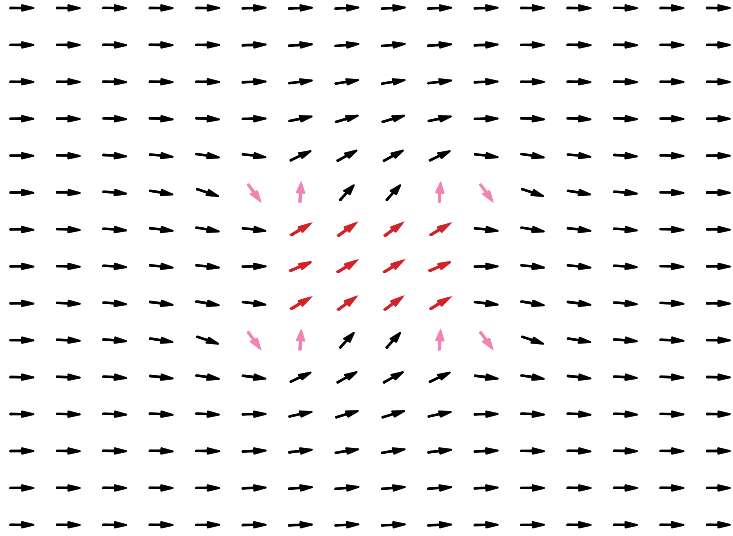,scale=0.75}
\caption{The ground state spin texture of the $4 \times 4$ square cluster
for four frustrating bonds. The pink spins are bond sites at which
the frustrating bonds are placed. For clarity, the spins within the
locally ordered cluster are coloured red.}
\label{fig:spins_gs}
\end{center}
\end{figure}

In the ground state, for which the spin distortion pattern is shown in 
Fig.~\ref{fig:spins_gs}, we have the opposite case: internally, the 
dipoles add constructively to give large distortions whereas outside 
they cancel to give very small ones.  The absence of long range spin 
distortions implies that these clusters can only weakly interact with 
other bonds. Most interesting, though, is that the internal spins are 
uniformly oriented but differently ordered from those spins outside the 
cluster.

\begin{figure}
\begin{center}
\epsfig{file=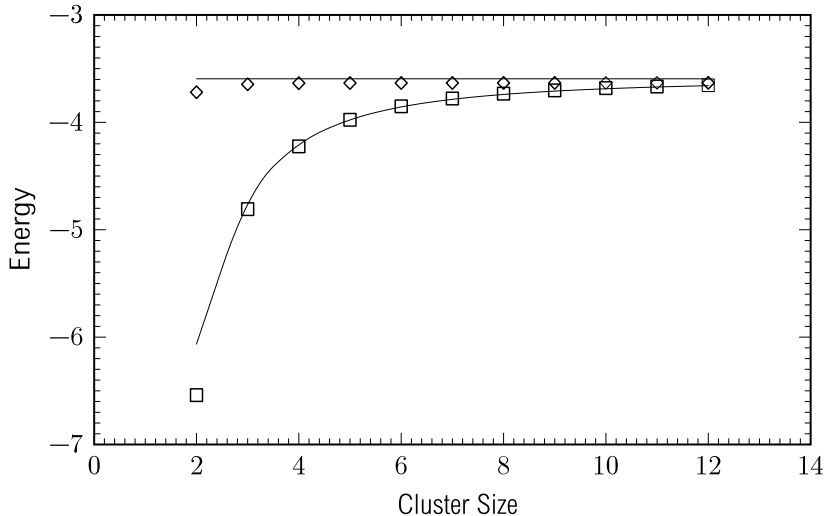,scale=0.75}
\caption{Energy of the ground (open squares and lower curve) and 
first excited (open diamonds and upper curve) states of the four 
frustrating bond arrangement described in the text. The 
open squares and diamonds are converged numerically generated
data points from simulations performed on the same size lattices 
as in Fig. 2.}
\label{fig:nrgs}
\end{center}
\end{figure}

Numerical solutions of the energies of this square cluster are shown in 
Fig.~\ref{fig:nrgs}. The solid curves are our analytical results (apart
from a constant times the single-bond energy (that is straightforward 
to calculate)), and provides strong support for their usage. 
In the ground state this figure makes clear that the binding
energy of the cluster can be quite large, especially for small cluster
sizes.  Further, square clusters tend to settle into states with
strong internal binding and which interact only weakly with other bonds.
That is to say, a square cluster is a locally ordered domain whose local
order parameter ${\bf \hat{\Omega}}$ is non-collinear with spins in the
rest of the lattice.  However, this behaviour is a strong function of
cluster size.  This construction, and the energy plot of 
Fig.~\ref{fig:nrgs}, demonstrates that such clusterings of spins 
(in regions that are not too large) are {\it stable}. 

The relation of such clusters to stripes, or in the case
of spin modulations, to the physics associated with
the appearance of domain walls, can be demonstrated by
considering the interface between such clusters, and to this end
we have analyzed a six-bond cluster given by
\begin{eqnarray}
{\bf r}^1  =  (0,0),~~
{\bf r}^2  &=&  (L,0),~~
{\bf r}^3  =  (2L,0),~~
{\bf r}^4  =  (0,L),~~
{\bf r}^5  =  (L,L),~~
{\bf r}^6  =  (2L,L),~~ \\ 
&&{\bf p}^{\alpha}  =  \pm p\hat{x} \ \ {\rm for}
~~\alpha \in C=\{1,2,3,4,5,6\}~~. \nonumber
\end{eqnarray}
Following the above analysis for four bonds, for the six-bond
situation there are 64 possible choices of the dipole moments'
orientations, and these states have 15 different energies. 
The lowest energy configuration corresponds to
\begin{equation} 
\left| \begin{array}{ccc} \rightarrow~~ \leftarrow~~ \rightarrow  \\
 \rightarrow~~ \leftarrow~~ \rightarrow \end{array} \right|
\end{equation}
and has a dipole-pair interaction energy of $(-103\pi J / 10)~(p/L)^2$
(which is noticeably lower than the first excited state,
which has an energy of $(-38\pi J / 10)~(p/L)^2$).

\begin{figure}
\begin{center}
\epsfig{file=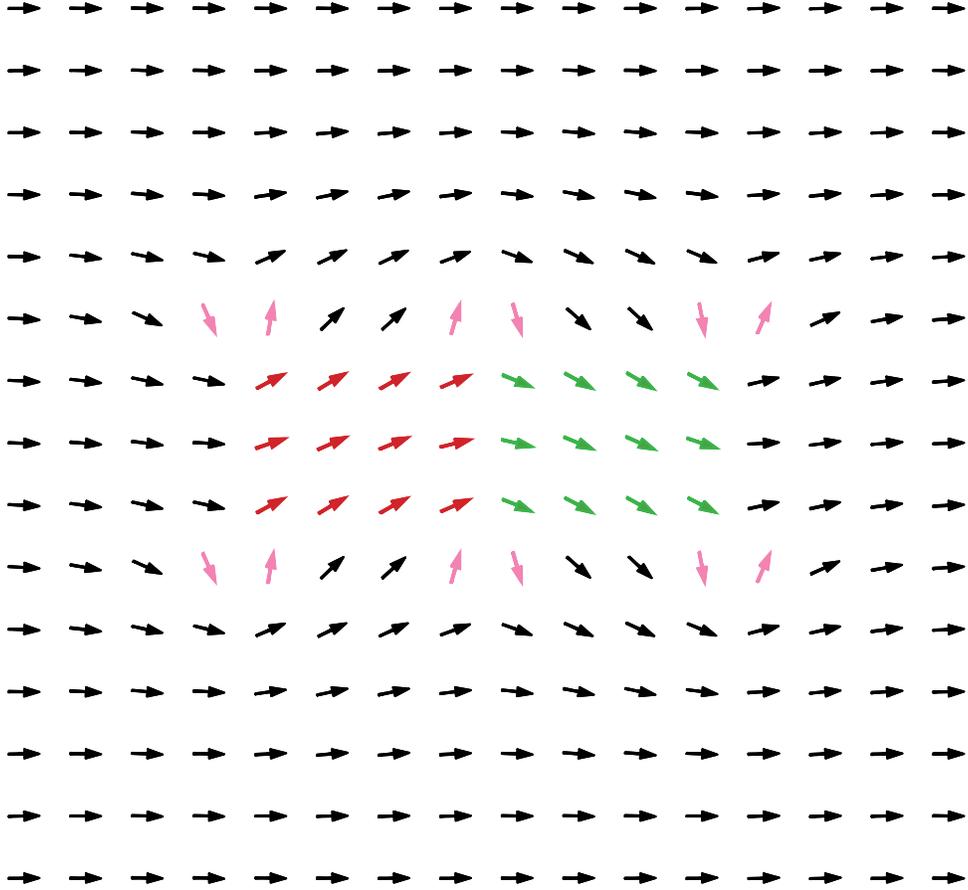}
\caption{Ground state of the six frustrating bonds showing the domain
wall between two clusters of locally ordered spins. As in Fig. 8,
the spins of the bond sites are coloured pink, while the two clusters
are coloured red and green, respectively.}
\label{fig:6bonds}
\end{center}
\end{figure}

The ground state spin texture for this location of the six
frustrating bonds is shown in Fig.~\ref{fig:6bonds}. From this
figure one can see the important result that this arrangement 
of spins is exactly what would expect if each $4 \times 4$ cluster 
within the 6-bond cluster was in its respective ground state.
Thus, between these $4 \times 4$ clusters one obtains a domain
wall, of width one lattice spacing, over which the local magnetic 
order parameter is rotated. Numerical evidence suggestive of
this type of domain wall was discussed at length in Ref. \cite {skyrms2}
for the case of a random distribution (and orientation) of
a non-zero density of frustrating spin interactions, and it is clear that
the same physics is at work in these two situations: spin twists.

\section{Conclusions:}

The above formalism has provided a detailed analytical theory of the 
spin distortions generated by frustrating bonds, and of the interactions
mediated by the spin background between them. We have critiqued its
validity by comparing to numerical solutions, and have found excellent
agreement. Then, by focusing on highly symmetrical distributions of 
frustrating bonds, reminiscent of local regions of bonds in the multiply 
doped state, we have used this theory to verify the existence of locally 
ordered magnetic clusters. Most importantly, our work shows that these 
clusters are stable. These are the clusters envisioned to exist in the 
so-called cluster spin-glass phase \cite {cho,skyrms2} of LSCO and 
${\rm Y_{1-x}Ca_xBa_2Cu_3)_6}$ \cite {Ca123}.

It is to be stressed that it is believed that mobile holes produce
the same kinds of spin distortions that are produced by the frustrating
bonds discussed in this paper \cite{ss1,frenkel}. Thus, we believe that 
our results support previous suggestions 
\cite{diluteFM,stojk,whitestripes} that spin twists and distortions are 
part of the competing interactions that might lead to rivers of charge 
appearing as low-energy fluctuations in the doped cuprate systems.

\acknowledgements

We wish to thank Marc-Henri Julien, John Tranquada, Kazu Yamada, 
Noha Salem, and especially Bob Birgeneau and David Johnston,
for helpful comments. Also, we thank Frank Marsiglio for a critical 
reading of the manuscript. The first draft of this paper was 
written while one of us (RJG) was visiting ICTP, Trieste, and he wishes 
to thank them for their hospitality and support. This work was supported 
in part by the NSERC of Canada.

\newpage
\appendix

\section{Single Bond Energy}

The question of the energy stored in the spin distortions can be answered by expanding
the Hamiltonian as follows:
\begin{eqnarray}
H & = & - J {\sum_{\langle ij \rangle}}' \cos \, (\phi_i - \phi_j) 
+ \lambda J \cos \, (\phi_1-\phi_0) \nonumber \\ 
& \approx & - J {\sum_{\langle ij \rangle}}' \left(1 - \frac{1}{2}
(\phi_i - \phi_j)^2 \right) + \lambda J \cos \, (2\phi_1) \nonumber \\ 
& = & - \sum_{\langle ij \rangle} J_{ij} + \frac{1}{2} J 
{\sum_{\langle ij \rangle}}' (\phi_i - \phi_j)^2 
- 2 \lambda J \sin^2 \, \phi_1
\end{eqnarray}

Of course, the term $-\sum J_{ij}$ represents the energy intrinsic to 
the lattice. Therefore, the energy from the spin distortions alone is 
given by
\begin{equation}
\frac{1}{2} J {\sum_{\langle ij \rangle}}' (\phi_i - \phi_j)^2 
               - 2 \lambda J \sin^2 \phi_1 \: .
\end{equation}

This in turn can be expanded using the continuum approximation
\begin{eqnarray}
E_{\mbox{\scriptsize dist}} 
& \approx & J \biggl\{ \frac{1}{2} \int_M (\nabla \phi)^2 d^2r 
+ (\phi_1 - \phi_4 )^2 
+ 2(\phi_1 - \phi_2 )^2 - 2 \lambda \sin^2 \phi_1 \biggr\} \nonumber \\ 
& = & J \biggl\{ \frac{1}{2} \int_{\partial M} \phi \nabla \phi \cdot {\bf ds} 
+(\phi_1 - \phi_4 )^2 + 2(\phi_1 - \phi_2 )^2 - 2 \lambda \sin^2 \phi_1 \biggr\}
\end{eqnarray}
where $M = {\cal R}\backslash D_{\epsilon}({\bf 0})$ is the $xy$-plane 
excluding a small region about the bond centre.
The integral term of this expression must be evaluated.  
Since the core of the bond occupies a unit disk at the origin, 
the appropriate value for $\epsilon$ is 1. Thus, one finds
\begin{equation}
E_{\mbox{\scriptsize dist}} = J \biggl\{ 
\frac{\pi}{2}\sum_k k A_k^2 + 
(\phi_1 - \phi_4 )^2 + 2(\phi_1 - \phi_2 )^2 
- 2 \lambda \sin^2 \phi_1 \biggr\} \:. 
\end{equation}
Using the results for the mode characterization of Eq.~(\ref{eq:ansatz1})
 gives
\begin{equation}
\sum_k k A_k^2 = \sum_k k {\left( \frac{2A_1}{k 2^k}\right)}^2 
= \frac{8}{\pi^2} \ln\left(\frac{5}{3}\right) \phi_1^2 
\end{equation}
and the identity $2\phi_2+\phi_4 = \phi_1$ allows us to write
\begin{equation}
(\phi_1-\phi_4)^2 + 2(\phi_1-\phi_2)^2 = \phi_1^2 + 2\phi_2^2+\phi_4^2 \:. 
\end{equation}
Thus the energy in the spin distortions is
\begin{equation}
E_{\mbox{\scriptsize dist}} = J \left\{ 
\left(\frac{4}{\pi}\ln\left(\frac{5}{3}\right) +1\right) \phi_1^2 + 
2\phi_2^2 + \phi_4^2 - 2 \lambda \sin^2 \phi_1 \right\} 
\end{equation}
which can be evaluated using Eq.~(\ref{eq:phi2_4 solns}) to give
\begin{equation}
E_{\mbox{\scriptsize dist}} = 2J \left( 
\phi_1^2 - \lambda \sin^2 \phi_1 \right) \: . 
\end{equation}
We stress that this expression includes the energy from the core
of the spin distortion field.

\section{Interaction Energy}

In accordance with Fig.~\ref{fig:bond_geom}, 
we write the spin distortion at the 0 site from the bond B as 
\begin{equation}
\phi_0^B  =  \frac{ {\bf p}^B \cdot ({\bf R} + \frac{1}{2}\hat{x})}
{|{\bf R} + \frac{1}{2}\hat{x}|^2} 
\approx  {\bf p}^B \cdot ({\bf R} + \frac{1}{2}\hat{x}) \frac{1}{R^2} 
\left[ 1 - \frac{{\bf R} \cdot \hat{x}}{R^2} - \frac{1}{4R^2} \right] \:.
\end{equation}
A similar expression can be written down for the other bond site.
Then, by superposition, the difference between the net distortions at 
sites 0 and 1 is 
\begin{equation}
(\phi_0 - \phi_1) =  (\phi_0^A - \phi_1^A) - \frac{2}{R^4}({\bf p}^B 
\cdot {\bf R}) ({\bf p}^B \cdot \hat{x}) 
+ \frac{1}{R^2}{\bf p}^B \cdot \hat{x} 
+ {\cal O}\left( \frac{1}{R^2} \right)
\end{equation}
Then we make use of the relationship between the spin distortions at the 
bond sites and the magnitude of the dipole moment:
\begin{equation} 
(\phi_0^A + \phi_1^A) \approx 2\frac{\pi}{2} p^A \:. 
\end{equation}
Therefore
\begin{equation} 
(\phi_0 - \phi_1)^2 = (\phi_0^A + \phi_1^A)^2 - \frac{8\alpha}{R^4}
({\bf p}^A \cdot {\bf R})({\bf p}^B \cdot {\bf R}) 
+ \frac{4\alpha}{R^2} {\bf p}^A \cdot {\bf p}^B 
\end{equation}
so that 
\begin{equation} 
\frac{1}{2}E = E({\bf p}^A) + \frac{1}{2}J\left\{
\frac{2\pi}{R^2} {\bf p}^A \cdot {\bf p}^B - \frac{4\pi}{R^4}
({\bf p}^A \cdot {\bf R})({\bf p}^B \cdot {\bf R}) \right\} 
\end{equation}
and thus, by symmetry, the total energy is
\begin{equation} 
E = E({\bf p}^A) + E({\bf p}^B) + J\{ \frac{2\pi}{R^2} {\bf p}^A 
\cdot {\bf p}^B - \frac{4\pi}{R^4}
({\bf p}^A \cdot {\bf R})({\bf p}^B \cdot {\bf R}) \} \:. 
\end{equation}
We conclude that the interaction energy is given by
\begin{equation}
E_{\mbox{\scriptsize int}}({\bf p}^A,{\bf p}^B) = 
J\frac{2\pi}{R^2}\left\{
\frac{2}{R^2}({\bf p}^A \cdot {\bf R})({\bf p}^B \cdot {\bf R})
-{\bf p}^A \cdot {\bf p}^B \right\} \:. 
\end{equation}
(Repeating the above calculation for two perpendicular
bonds produces the identical result.)


\newpage
\begin{references}

\bibitem[\dagger]{byline} Present Address: Dept. of Physics, 
Massachusetts Institute of Technology, Cambridge, MA 02139.

\bibitem{jtran review} For a recent comprehensive review, see 
J. Tranquada, in ``Neutron Scattering in Layered Copper-Oxide 
Superconductors'', edited by A. Furrer (Kluwer, Dordrecht, The Netherlands,
1998), pp. 225; also, see references therein.

\bibitem{ek1} V.J. Emery and S.A. Kivelson, 
Phys. Rev. Lett. {\bf 64}, 475 (1990).

\bibitem{whitedec99} For a recent discussion of this issue, see
 S. Rommer, S.R. White, and D.J. Scalapino, cond-mat/9912352.

\bibitem{fps} V.J. Emery and S.A. Kivelson, 
Physica C {\bf 209}, 597 (1993).

\bibitem{ames} F. Borsa, P. Carreta, J.H. Cho, F.C. Chou, Q. Hu, 
D.C. Johnston, A. Lascialfari, D.R. Torgeson, R.J.  Gooding, N.M. Salem, 
and K.J.E. Vos, Phys. Rev. B {\bf 52}, 7334 (1995).

\bibitem{kyamada} K. Yamada, ICTP, Trieste, July 12-23, 1999, 
and private communication.

\bibitem{lai} E. Lai and R.J. Gooding, Phys. Rev. B {\bf 57}, 1498 (1998).

\bibitem{cho} J.H. Cho and F. Borsa and D.C. Johnston and D.R. Torgeson,
Phys. Rev. B, {\bf 46}, 3179 (1992).

\bibitem{skyrms2} R.J. Gooding, N.M. Salem, R.J. Birgeneau and F.C. Chou, 
Phys. Rev. B, {\bf 55}, 6360 (1997).

\bibitem{conun} Of course, one way around this 
disagreement is that there could be more than one way to produce a 
cluster spin-glass phase. 

\bibitem{mhj} M.-H. Julien, {\em et al.},
Phys. Rev. Lett. {\bf 83}, 604 (1999).

\bibitem{ss1} B.I. Shraiman and E.D. Siggia, 
Phys. Rev. Lett. {\bf 61}, 467 (1988). 

\bibitem{diluteFM} N.M. Salem and R.J. Gooding, Europhys. Lett.
{\bf 35}, 603 (1996).

\bibitem{stojk} B. Stojkovic, {\em et al.}, cond-mat/9805367;
{\em ibid}, cond-mat/9911380.

\bibitem{whitestripes} See the discussion in S.R. White and 
D.J. Scalapino, cond-mat/9907243, and references therein.

\bibitem{aharony} A. Aharony, {\it et~al.}, Phys. Rev. Lett. {\bf 60}, 
1330 (1988).

\bibitem{skyrms1} R.J. Gooding, Phys. Rev. Lett. {\bf 66}, 2266 (1991); 
R.J. Gooding and A. Mailhot, Phys. Rev. B {\bf 48}, 6132 (1993).

\bibitem{doping} Note that $x/2$ is the probability of a bond having an 
oxygen hole present, thus implying that $x$ is the probability of any 
plaquette having one hole present, as per usual.

\bibitem{Ca123} Ch. Niedermayer, {\it et~al.}, 
Phys. Rev. Lett. {\bf 80}, 3843 (1998).

\bibitem{vann} J. Vannimenus and S. Kirkpatrick and F. D. M. Haldane 
and C. Jayaprakash, Phys. Rev. B, {\bf 39}, 4634 (1989).

\bibitem{higherorder} The expansion of the continuum limit of
the equilibrium condition can be generalized to higher order.
Nonetheless, since we expect $\phi$ to vary slowly as a 
function of position away from the bond, the second order equation 
({\em viz.}, Laplace's equation) shall suffice. If one does examine
the higher order terms one finds higher angular harmonics for each
order of $1/r^m$ which are small perturbations to the exact solution.

\bibitem{syms} As per usual, it is useful to consider what symmetries 
such a solution might exhibit. The natural point symmetry group on the 
undoped lattice is the dihedral group of order 8, $D_8$. An AFM bond 
introduces a preferred direction in space and immediately singles out 
two lines, one running through the bond and the other bisecting it, 
as special. Suppose that the bond is $x$-directed and is positioned 
suitably far from the edges of a very large lattice. Then, using a 
coordinate system centred on the bond, we can say that the physical 
situation possesses reflectional symmetry across the $x$- and $y$-axes; 
denote its symmetry group by $G \subset D_8$. We expect solutions 
$\phi_i$ to the spin distortions to share these same symmetries in the 
sense that $\phi_i$ is invariant (up to a global sign change) under 
transformation by any $g \in G$. That is, at all sites $i$, either
$\phi_i = +(\phi \circ g)_i$ or $\phi_i = -(\phi \circ g)_i$. 
In fact, we find that ground state solutions are odd in reflection 
across the $y$-axis and even in reflection across the $x$-axis; 
this information will be employed in the text.  

\bibitem{bogdan} A. S. Kovalev and M. M. Bogdan, 
Sov. Solid State Phys., {\bf 35}, 886 (1993).

\bibitem{aharonynmr} C. Goldenberg and A. Aharony, 
Phys. Rev. B {\bf 56}, 661 (1997).

\bibitem{grempel} P. Gawiec and D. R. Grempel, 
Phys. Rev. B, {\bf 44}, 2613 (1991).

\bibitem{frenkel} D.M. Frenkel, R.J. Gooding, B.I. Shraiman, 
and E.D. Siggia, Phys. Rev. B {\bf 41}, 350 (1990).

\end{references}
\end{document}